\documentclass[aps,pra,10pt,twocolumn,showpacs,superscriptaddress,footnoteinbib]{revtex4}
\usepackage{amsmath}
\usepackage{latexsym}
\usepackage{amssymb}
\usepackage{color}
\usepackage{graphics,epstopdf}
\usepackage{soul}

\begin{document}

\title{Classical light steering leading to quantum-like security}

\author{Tanumoy Pramanik}
\email{tanu.pram99@bose.res.in}
\affiliation{S. N. Bose National Centre for Basic Sciences, Block JD, Sector III, Salt Lake, Kolkata 700098, India}

\author{A. S. Majumdar}
\email{archan@bose.res.in}
\affiliation{S. N. Bose National Centre for Basic Sciences, Block JD, Sector III, Salt Lake, Kolkata 700098, India}

\date{\today}

\begin{abstract}

We show how  single system steering can be  exhibited by classical light, a feature 
originating from superposition in classical optics that also enables entanglement and 
Bell-violation by classical light beams. Single system steering is the temporal
analogue of Einstein-Podolsky-Rosen (EPR) steering in the quantum domain, 
enabling control of the state of a remote system, and can hence be connected to 
the security of secret key generation between two remote parties.
We derive the steering criterion for a single mode coherent state when displaced parity 
measurements are performed at two different times. The security bound of the 
Bennett-Brassard 1984 (BB84) protocol under the gaussian cloning attack is calculated 
to yield an, in principle, ideal and quantum-like key rate using a fine-grained
uncertainty relation corresponding to the classical phase space.

\end{abstract}

\pacs{03.67.-a, 03.67.Mn}

\maketitle


 In the quantum world the superposition principle plays a fundamental role in 
nearly all phenomena, without which it would be impossible for entanglement to exist
and be used as resource for performing quantum information processing tasks.
Among the many examples of quantum information processing, quantum key distribution
(QKD) is one of the most widely studied~\citep{reviews, BB84, E91, DIQKD, 1sdiqkd}, in 
view of its importance in the practical 
demonstration of the quantum advantage over corresponding classical information processing
protocols. The security of QKD protocols is guaranteed, in principle, by quantum
uncertainty~\cite{BB84}, and is further linked in entanglement based protocols to 
quantum nonlocality~\citep{E91, DIQKD} and quantum steering~\citep{1sdiqkd}. The first QKD
protocol to be proposed, {\it viz.} the BB84 protocol is based on the superposition
principle with its security ensured by the uncertainty relations.

The relevance of superposition in physics, is however, not exclusive to the quantum domain.
The importance of superposition in classical wave theory
was manifested through interference and diffraction phenomena long before the advent of 
quantum 
theory. In modern times classical optical coherence in utilized 
in wide ranging applications such as
holographic interferometry~\cite{holog} and magneto-optical Kerr effect~\cite{kerr} in the
study of structure of materials, to interferometric telescopy and Hanbury Brown-Twiss 
effect~\cite{hanbury} in astronomy. Classical light beams with non-trivial topological 
structure 
have been discovered~\cite{berry} with applications in optical tweezers~\cite{tweez}, and
are regarded to be potentially useful for information processing due to their ability to
carry large amounts of information~\cite{cinfo}. 
     
The key role of superposition common to both quantum mechanics and classical optics has lead
to the formulation of uncertainty relations in the latter analogously to the well-known
uncertainty principle of the former~\cite{C-Op, mansuri}. In wave optics the wavelength
of light $\lambdabar=\lambda/2\pi=c/\omega$ plays a role analogous to the Planck's constant 
$\hbar$ in quantum mechanics. The finite and non-vanishing wavelength $\lambdabar$ leads to
the lack of precision in simultaneous measurement of two incompatible observables. In other
words, relationships analogous to the Heisenberg uncertainty relation, such as
\begin{eqnarray}
(\Delta \hat{x}^2)(\Delta \hat{p}^2) \ge \frac{\lambdabar^2}{4}
\label{classuncert}
\end{eqnarray}
 are obtained
between observables corresponding to the position space and the wave vector space due to
the finite and non-vanishing wave vector $k \sim 1/lambda$ involved in fourier 
transformation connecting these two domains. The above analogy results in mathematical
isomorphism for correlations between physical degrees of freedom in classical
optics in relation with quantum entanglement in two-qubit systems, thus inspiring the
formulation of the theory of classical entanglement~\cite{classent} and violation of Bell 
inequalities~\cite{BCHSH} in classical electromagnetism~\cite{classbell, priyanka}. In 
particular, Schmidt
decomposition pertaining to superposition of classical electromagnetic fields~\cite{agar1} 
has been exploited to derive Bell-like inequalities~\cite{priyanka} for classical 
vortex beams~\cite{agarbook}. Such Bell violation in the domain of classical continuous
variable phase space has been experimentally verified too~\cite{rpsingh}.

Other than entanglement and Bell nonlocality, another form of correlation in quantum mechanics
is exemplified by EPR steering~\cite{epr}. Steering entails the ability to control the
state of a remote system through local measurements. Formulation of correlations in
terms of their applicability in
information theoretic tasks involving two distant parties enables understanding of
quantum steering as an intermediate correlation between entanglement and Bell 
nonlocality~\cite{wise}. Besides the above types of spatial correlations, the quantum
framework accommodates certain temporal correlations, such as those quantified by the 
Leggett-Garg inequality~\cite{lgineq}, as well as temporal steering or single system 
steering~\cite{St_Tem_Dis}. Inspired by the above analogy in features of classical optics
and quantum mechanics, such as superposition, entanglement and Bell violation, we are thus
motivated to enquire as to what task analogous to quantum steering may be implementable in
classical optics.     

In the present work we develop a protocol for single system steering in classical optics.
Quantum steering of single systems has been formulated recently~\cite{St_Tem_Dis}, and
shown to have applications in the security of the BB84 key distribution protocol, as well
as in quantifying non-Markovianity~\cite{nonmarkov}. All steerability conditions arise
from uncertainty relations which form the underlying basis of security in key distribution
protocols. EPR steering has been linked directly to the key rate of one-sided device
independent key distribution~\cite{1sdiqkd}, and similarly, single system steering to the
BB84 key distribution~\cite{St_Tem_Dis}. Optimal steering relations have been 
derived~\cite{St_F_D} using the fine-grained uncertainty relation that connects uncertainty
with nonlocality in quantum mechanics~\cite{FUR_D}. A continuous variable 
fine-grained steering inequality has also been derived which leads to an, in principle,
ideal key rate in one-sided device independent key distribution~\cite{St_F_C}. Here,
in order to derive an optimal steering inequality in classical electromagnetism, we first
formulate a fine-grained uncertainty relation in classical phase space. Our
fine-grained steering condition thus forms the basis for the single systems steering
protocol in classical optics, and is finally used to obtain an, in principle, ideal key 
rate for the BB84 protocol using a single mode coherent state.

We begin by discussing briefly the key features of uncertainty in the phase space of
classical optics~\cite{C-Op, mansuri, priyanka}. In paraxial optics the  propagation of a 
light beam 
$E(\vec{r},t)=\varepsilon(\vec{r})\,\left(\frac{i\omega z}{c}-i\omega t\right)$ in free 
space is described by 
$i \frac{\partial \varepsilon}{\partial z} = -\frac{\lambdabar}{2} \nabla_k^2 \varepsilon$,
(with $\lambdabar=\lambda/2\pi=c/\omega$ and $k=x,y$), which is, for $t\rightarrow z,\;\psi\rightarrow\varepsilon,\;\hbar\rightarrow\lambdabar$, exactly the Schrodinger equation for a free 
particle in two dimensions.  Here $\lambdabar \to 0$
leads to the limit of geometrical optics in a similar way as
 $\hbar \to 0$ yields the classical limit of quantum mechanics.  Therefore, any optical 
beam in two dimensions can be written as 
superposition of the solutions of the above equation.  For example, eigenfunctions of the two
dimensional harmonic oscillator can be expressed as Laguerre-Gaussian beams constructed from 
the superposition of Hermite-Gaussian functions~\cite{agarbook}. Exploiting this feature,
classical entanglement and Bell violation has been demonstrated for Laguerre-Gaussian 
beams~\cite{priyanka}. The quadrature amplitudes corresponding to the field 
$E_{\alpha_j} \propto \hat{\alpha}\exp^{-\omega_{\alpha_j}t} + \hat{\alpha}^{\dagger}\exp^{i\omega_{\alpha_j}t}$
(with the two modes denoted by $a_j$ $j=1,2$) may be written as 
$\hat{X}_i^{\theta_j} = \frac{ \hat{a}_i\exp\left(-i\theta_i\right) + \hat{a}_i^{\dagger}\exp\left(i\theta_i\right)}{\sqrt{2}}$,
where $\hat{a}_i =  \frac{X_i + iP_i}{\sqrt{2}}$, in terms of the dimensionless variables $X_i$
and $P_i$ defined as $X_i = \frac{\sqrt{2} x_i}{\sigma}$ and $P_i = \frac{\sigma p_i}{\sqrt{2} \lambdabar} $, with $\sigma$ being any suitable parameter of length dimensions, for instance,
the beam waist in the case of Laguerre-Gaussian 
beams~\cite{priyanka}, or 
the initial position space width of Gaussian beams. Therefore, the commutation relations $\left[\hat{x}_i, \hat{p}_j\right] = i\lambdabar \delta_{ij}$
become $\left[\hat{X}_i, \hat{P}_j\right] = i\delta_{ij}$, with $\hat{P}_i = -i\frac{\partial}{\partial X_i}$, leading to $\left[\hat{a}_i, \hat{a}_j\right] = \delta_{ij}$. Thus, the 
  Heisenberg type uncertainty relation for the dimensionless phase space variables $X$ and
$P_X$,  is obtained from $(\Delta {X}^2) ~ (\Delta {P}_x^2) \ge \langle[X, P_X]\rangle^2$, and is given by
\begin{eqnarray}
(\Delta {X}^2) ~ (\Delta {P}_x^2)\geq \frac{1}{4}.
\label{HUR2}
\end{eqnarray}

Limitations of the Heisenberg uncertainty relation in quantum mechanics~\cite{HUR} were 
noted soon after 
its formulation by  Schrodinger and Robertson~\cite{GUR} who presented an improved version
 for any two arbitrary observables. Subsequently, a number of works were performed to 
alleviate inadequacies such as the state dependence of the lower bound of uncertainty, as
well as to develop the uncertainty 
relation for information theoretic purposes~(see the reviews~\cite{UR_Survey, UR_Rev}).  
 An entropic uncertainty relation (where uncertainty is 
measured by Shannon entropy) was derived in wave mechanics~\cite{EUR-C} with the help of 
the $(p,q)-$ norm of the 
Fourier transformation. It is now 
accepted that to 
characterize different tasks in quantum information theory, entropy as a measure of 
uncertainty is more useful than standard deviation\cite{St_C}. For example, entropic 
uncertainty relations
show quantum steering by certain continuous variable non-Gaussian states~\cite{LG_Steer}, which is
failed to be revealed by the Heisenberg uncertainty relation.
The entropic uncertainty relation corresponding to our 
classical wave mechanical phase space directly follows from the work of Bialynicki-Birula and 
Mycielski~\cite{EUR-C}, and is given by
\begin{eqnarray}
\mathcal{H}({X}) + \mathcal{H}({P}_X) \geq \ln(\pi e),
\label{Eur_C}
\end{eqnarray}
where $\mathcal{H}({X})=\int dX |\Psi(X)|^2 \ln |\Psi(X)|^2 $ and $\mathcal{H}({P}_X)=\int dk_X |\Psi(k_X)|^2\ln|\Psi(k_X)|^2$, with $\Psi(X)$  and $\Psi(k_X)$  the position space and
wave vector space wave functions, respectively.  The entropic uncertainty relation~(\ref{Eur_C}) 
implies the Heisenberg uncertainty relation~(\ref{HUR2})~\cite{EUR-C}, since the two 
uncertainty relations (\ref{HUR2}) and (\ref{Eur_C}) are connected by the inequality
$-\mathcal{H}(\alpha) \leq \frac{1}{2} \ln(2\pi e \Delta \alpha^2)$,
where $\alpha\in\{{X}, {P}_X\}$.

It has been recently realized that
it is possible to construct several quantum games where coarse grained uncertainty relations, 
such as the Heisenberg and entropic uncertainty relations (where uncertainty is measured in 
a coarse grained way by taking the  average of uncertainty over all possible measurement 
outcomes) fail to give optimal playoff~\cite{FUR_D, Q_memory}. Fine-grained forms of uncertainty 
relations are able to handle such situations better by quantifying uncertainty directly in terms of
 probabilities of getting particular measurement outcomes and their combinations. Fine-graining
reveals the connection of uncertainty with nonlocality of the underlying physical theory for
bipartite and tripartite qubit quantum games~\cite{FUR_D, tanu1}. Fine-grained uncertainty
relations have been further used to derive optimal steerability conditions for both 
discrete~\cite{St_F_D}  and continuous~\cite{St_F_C} variable systems. The fine-grained uncertainty
relation relevant to the phase space considered here may be obtained with the help of 
displaced parity 
measurement of  $\Pi(\beta)=\Pi^+(\beta) - \Pi^-(\beta)$,
where $\Pi^+(\beta)=\mathcal{D}(\beta)\sum_{n=0}^{\infty}|2n\rangle\langle 2n|\mathcal{D}^{\dagger}(\beta)$,  and $\Pi^-(\beta)=\mathcal{D}(\beta)\sum_{n=0}^{\infty}|2n+1\rangle\langle 2n+1|\mathcal{D}^{\dagger}(\beta)$ correspond to even and odd parity operators, respectively,  with $D(\beta)=\exp\left[\beta b^\dagger - \beta^* b\right] $ being the displacement operator. $\Pi(\beta)$ and $\Pi(-\beta)$ are associated
 with uncertainty 
relations~\cite{St_F_C}. The displacements $``+\beta"$ and 
$``-\beta"$ are chosen with the probabilities $P_{\beta}$ and $P_{-\beta}$, respectively,
with $P_{\beta} + P_{-\beta}= 1$.    Labelling the probabilities for getting odd (even) parity measurement
outcomes by $b=0$ ($b=1$), the probability distribution $\left[P_{\beta} P(b_{\beta})\,+\,P_{-\beta}P(b_{-\beta})\right]$
is bounded by~\cite{St_F_C}
\begin{eqnarray}
\frac{1}{4}\leq \left[P_{\beta} P(b_{\beta}=0)\,+\,P_{-\beta}P(b_{-\beta}=0)\right]\leq \frac{3}{4},
\label{FUR_C0}
\end{eqnarray}
In order to see the connection~\cite{UR_Rev} between the entropic and fine-grained forms of 
uncertainty relations, consider the  R\'{e}nyi entropy of order $\eta$  given by
$\mathcal{H}_{\eta}=\frac{1}{1-\eta}\log\left(\sum_{b=1}^n p^{\alpha}(b)\right)$.
Shannon entropy is the R\'{e}nyi entropy with order $\eta\rightarrow1$, while Min-entropy is 
the R\'{e}nyi entropy with order $\eta\rightarrow \infty$  defined by
$H_{\infty}=\min_{b}\left[-\log P(b)\right]=-\log\max_b P(b)$. Now, setting $P_{\beta} = 1/2 = P_{-\beta}$
for simplicity, and using the concavity of the $\log$ function, it follows that
$\frac{1}{2} \mathcal{H}_{\infty}(\beta) + \frac{1}{2} \mathcal{H}_{\infty}(-\beta) \geq  -\log\max\left[\frac{1}{2} P(b_{\beta})\,+\frac{1}{2}\,P(b_{-\beta})\right]$. Next, using the second inequality
in Eq.(\ref{FUR_C0}), one gets $\mathcal{H}_{\infty}(\beta) +  \mathcal{H}_{\infty}(-\beta) \geq  -2\log\frac{3}{4}$, which is of the form similar to the entropic uncertainty relation given by
Eq.(\ref{Eur_C}).


Using the above fine-grained uncertainty relation (\ref{FUR_C0}), we are now equipped to present our 
protocol for steering in classical wave optics.  In the usual EPR steering 
scenario~\cite{wise, St_C, St_Exp1}, Alice prepares systems $A$ and $B$ in the  bipartite quantum 
state $\rho_{AB}$ and sends the system $B$ to Bob. Bob accepts that the shared state $\rho_{AB}$ is 
steerable, only when Alice can control the state of the system $B$, which is demonstrated by the
violation of a suitable steering inequality based upon the corresponding uncertainty relation. 
However, in the temporal steering scenario~\cite{St_Tem_Dis, nonmarkov}, Alice prepares a single 
quantum systems $A$ in a well known state (by measuring the system in certain basis) $\sigma_A$ at 
time $t_1$ and sends the system to Bob who checks her control over the state of that system  at 
a later time $t_2$. If no noise effects such as environmental decoherence, or eavesdropping occurs
in the elapsed interval ($t_2 - t_1$), Alice has complete information about the state of $A$, 
and she can thus project the state  in any basis by communicating a suitable unitary rotation to 
Bob. Temporal steering conditions are mathematically similar to the EPR steering inequalities.

The temporal steering criterion relevant to our present work amy be obtained by considering the 
following game. Alice first prepares a large number of identical systems labeled by $A$ in single 
mode coherent state given by\footnote{Note that though we have earlier discussed the analogy between 
quantum  and classical wave mechanics in the context of two-dimensional paraxial optics, 
for our present purpose correlations among two modes as in the case of classical entanglement and 
Bell-violation~\cite{priyanka} are not required here. At a formal level, the two-dimensional 
position space wave function may, for example,  be taken to be a product of two Gaussians with the 
uncertainty relations (\ref{HUR2}), (\ref{Eur_C}) and (\ref{FUR_C0}) in classical theory  obeyed 
independently by the two corresponding uncoupled modes. It thus suffices to consider a single mode 
system for the rest of the analysis presented in this work.} 
\begin{eqnarray}
|\alpha\rangle_A=\exp\left[-\frac{|\alpha|^2}{2}\,\sum_{n=0}^{\infty}\frac{\alpha^n}{\sqrt{n!}}|n\rangle_A\right],
\label{Coherent}
\end{eqnarray}
 Next, at time $t_1$, she randomly applies either the displacement operator $\mathcal{D}(\beta)$ or $\mathcal{D}(-\beta)$, where,  for simplicity, we consider $\beta$ to be real. Therefore, the system 
$A$ is prepared in the states $|\alpha + \beta\rangle_A$ and $|\alpha - \beta\rangle_A$  with the 
probabilities  $P_{\alpha+\beta}$ and
$P_{\alpha-\beta}$, respectively.  The probability of getting parity measurement outcome $a$ ($a=0$ for even parity and $a=1$ for odd parity) for the state chosen from the set $\{|\alpha+\beta\rangle,\,|\alpha-\beta\rangle\}$ is bounded by the uncertainty relation~(\ref{FUR_C0})
\begin{eqnarray}
\frac{1}{4}\leq \left[P_{\alpha+\beta} P(a_{\alpha+\beta})\,+\,P_{\alpha-\beta}P(a_{\alpha-\beta})\right]\leq \frac{3}{4},
\label{FUR_C}
\end{eqnarray}
where  $P(a_{\alpha\pm\beta})$ is the probability of obtaining the outcome $a$ for the state 
$|\alpha\pm\beta\rangle$, and the values $\{\alpha\rightarrow 0 ~ \& ~ \beta\rightarrow 0\}$
are excluded~\cite{St_F_C}.  After this Alice sends the system $A$ to Bob who 
does not have any prior knowledge of the state of the system. Alice then communicates to Bob over 
a public channel informing him to apply the displacement operator $\mathcal{D}(\gamma_1)$ when the 
prepared state is $|\alpha+\beta\rangle$, and $\mathcal{D}(\gamma_2)$ for the prepared state 
$|\alpha-\beta\rangle$. To check Alice's steerability,   Bob measures at time $t_2$ the parity 
of the displaced set of states $\{\mathcal{D}(\gamma_1)|\alpha+\beta\rangle,~ \mathcal{D}(\gamma_2)|\alpha-\beta\rangle\}$.

For noiseless channels shared between Alice and Bob, the state sent by Alice is the same as the 
state received by Bob. Therefore, Alice has full control over Bob's system. In practice, the 
channels are not ideal due to environmental interaction or interruption by a third party. Noise
 introduces unexpected uncertainty in the conditional probability of Bob's 
measurement outcome given  Alice's preparation. The (non-)steerability condition may be 
derived by assuming  that  noise
makes the states $|\alpha\pm\beta\rangle$ end up in an unknown state $\sigma_\lambda$ with 
the conditional probability $P(\lambda|\alpha\pm\beta)$. Bob thus receives the  system $A$ in the 
state $\sigma_B(\alpha\pm\beta)
=\sum_{k=0}^1\sum_{\lambda} P\left(\alpha+(-1)^k\beta\right)P\left(\lambda|\alpha+(-1)^k\beta\right)\sigma_\lambda
= \sum_\lambda P(\lambda) \sigma_{\lambda}$,
when the prepared state is $|\alpha\pm\beta\rangle$. It is clear that Bob's state $\sigma_B$ is 
independent of Alice's preparation procedure. Hence, Alice does not have any control over the 
state $\sigma_B$, i.e., the state is unsteerable. The sum of probabilities obtained
using the state $\sigma_B(\alpha\pm\beta)$  and also the 
state $\mathcal{D}(\gamma_{1})\sigma_B(\alpha\pm\beta)\mathcal{D}^{\dagger}(\gamma_{1})$  satisfy the 
inequality~(\ref{FUR_C}).

On the other hand, if Alice has control over Bob's system, Bob can reduce his uncertainty about 
measurement outcomes when Alice sends the information about the displacement operator. The sum of 
the conditional probabilities of getting 
the outcome $b$ for the parity measurement after applying the 
displacement operator $\mathcal{D}(\gamma_1)$ by Bob  for the prepared state 
$|\alpha+\beta\rangle_A$  and for the case when Bob applies the displacement operator 
$\mathcal{D}(\gamma_2)$  for the prepared state $|\alpha-\beta\rangle_A$, i.e., $\left(P_{\alpha+\beta}P(b_{\gamma_1}|\alpha+\beta)+P_{\alpha-\beta}P(b_{\gamma_2}|\alpha-\beta)\right)$ will lie outside of the region $[1/4,\,3/4]$ 
in this case.   Therefore,  temporal steerability occurs when either of the two
inequalities
\begin{eqnarray}
\frac{1}{4} \le  \left[P_{\alpha+\beta}P(b_{\gamma_1}|\alpha+\beta)+P_{\alpha-\beta}P(b_{\gamma_2}|\alpha-\beta)\right] 
\le \frac{3}{4},
\label{TSC_U}
\end{eqnarray}
is violated. For example, with the choice of displacements $\gamma_1 = -\alpha -\beta$ and
$\gamma_2 = -\alpha +\beta$, the sum of probabilities is plotted versus $\beta$ and $P_{\alpha+\beta}$
in the Figure 1, showing that the state (\ref{Coherent}) is steerable in the ranges
$0<p<\frac{\sqrt{2} \sqrt{2 e^{-8\beta ^2}-3 e^{-4 \beta^2}+1}+2 e^{-4 \beta^2}-2}{4 \left(e^{-4 \beta^2}-1\right)}$ and $\frac{-\sqrt{2} \sqrt{2 e^{-8 \beta ^2}-3 e^{-4 \beta^2}+1}+2 e^{-4 \beta^2}-2}{4 \left(e^{-4 \beta^2}-1\right)}<p<1$.
Specifically, if Alice has full control over Bob's system, the probability of 
getting  even parity is unity for $\gamma_1=-\alpha -\beta$ and $\gamma_2=-\alpha+\beta$, and 
similarly, the probability of getting odd parity is unity for the choice $\gamma_1=1-\alpha-\beta$ 
and $\gamma_2=1-\alpha+\beta$.
Note that the above steerability consition is obtained in the context
of classical optics, and is based on the fine-grained uncertainty relation~(\ref{FUR_C0})
for classical phase space variables. One may obtain similar steerability conditions
in classical optics based on other forms of uncertainty relations such as those
given by Eqs.({\ref{HUR2})
and (\ref{Eur_C}). Here we employ the fine-grained form of uncertainty since it provides an
optimal steerability condition for continuous variables~\cite{St_F_C}.

\begin{figure}[!ht]
\resizebox{9cm}{6cm}{\includegraphics{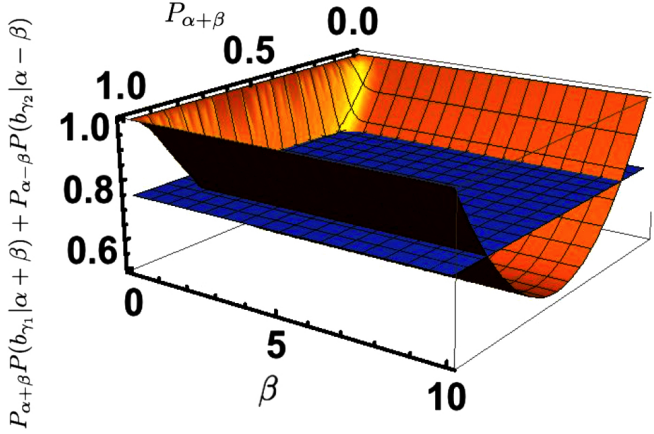}}
\caption{\footnotesize Steerability of the state (\ref{Coherent}) for the choice $\gamma_1=-\alpha -\beta$ and $\gamma_2=-\alpha+\beta$. The  curved and plane surfaces represent the sum of probabilities
and the upper bound, respectively, in
Eq.(\ref{TSC_U}).
\label{Fig_Ex}
}
\end{figure}

Temporal steering is connected to the security of the BB84 protocol~\cite{BB84} 
of key generation, as was shown for the case of discrete variables~\cite{St_Tem_Dis} in
quantum mechanics.
 Here we discuss the security  of a key generation scenario in classical optics similar to
the BB84 protocol using continuous variables. Here, Alice prepares an ensemble of systems $A$ 
either in the coherent state $|\alpha +\beta\rangle$ or in $|\alpha-\beta\rangle$.  When Alice 
sends the systems $A$ to Bob, an eavesdropper, Eve can clone the system to gain information about 
it.  Perfect cloning is disallowed in quantum mechanics as it would enable determination of
two incompatible observables through an arbitrarily large number of cloned copies. 
Perfect cloning is impossible in the 
phase space of classical optics in order to preserve the uncertainty relation~(\ref{HUR2}),
analogously to the case of quantum mechanics where perfect cloning is forbidden in order to
preserve the quantum uncertainty principle~\cite{noclone}.  In continuous variable 
systems the optimal strategy of cloning is achieved using the Gaussian cloning machine~\cite{GQC} 
by which coherent states are cloned with the optimal fidelity $2/3$ .
The operation of the Gaussian cloning on the state $|\alpha\pm\beta\rangle_A$ leads to
\begin{eqnarray}
|\alpha\pm\beta\rangle_A\,|0\rangle_E \rightarrow |\left(\alpha\pm\beta\right)\cos|\eta|\rangle_A|\left(\alpha\pm\beta\right)\frac{\eta}{|\eta|}\,\sin|\eta|\rangle_E,
\label{G_Cloning}
\end{eqnarray}
where $|0\rangle_E$ the initial state of Eve's system  and $\eta$ is the cloning parameter. 
Considering application of the displacement operators $\gamma_1=-\alpha-\beta$ or
$\gamma_2=-\alpha+\beta$, due  to Eve's interception at the time of its transit, the final state of the system $A$ becomes  $|\left(\alpha+\beta\right)\left(\cos|\eta|-1\right)\rangle_A$ or $|\left(\alpha-\beta\right)\left(\cos|\eta|-1\right)\rangle_A$. When Alice tries to create an even ($0$) parity state on
Bob's side, she informs him of the choice of displacement $\gamma_1=-\alpha-\beta$ for her prepared
state $\alpha+\beta\rangle_A$, or $\gamma_2=-\alpha+\beta$ for her prepared state $\alpha-\beta\rangle_A$. However, error occurs due to eavedropping, and hence, the probability of getting odd parity ($1$) by 
Bob becomes
\begin{eqnarray}
P_{01}(\alpha\pm\beta)&=&\sum_{m=0}^{\infty} |\langle 2m+1|\left(\alpha\pm\beta\right)\left(\cos|\eta|-1\right)\rangle_A|^2 \nonumber \\
&=& \sinh\left(|\delta|^2\right)\,\exp\left(-\frac{|\delta|^2}{2}\right),
\end{eqnarray}
where $\delta=\left(\alpha\pm\beta\right)\left(\cos|\eta|-1\right)$, and for simplicity we take
$P_{\alpha+\beta} = 1/2 = P_{\alpha-\beta}$.  As Alice randomly prepares the system $A$ either in the state $|\alpha+\beta\rangle_A$ or in the state $|\alpha-\beta\rangle$. The average error is given by
$P_{01}=\frac{1}{2} P_{01}(\alpha+\beta) + \frac{1}{2} P_{01}(\alpha-\beta)$.
The correlation between Alice and Bob is quantified by mutual information, $\mathcal{I}(A:B)$ which is defined by
$\mathcal{I}(A:B)= \mathcal{H}(A) -\mathcal{H}(B|A)$,
where $\mathcal{H}$ is Shannon entropy. As Alice randomly prepares system A in the state $|\alpha\pm\beta\rangle_A$, $H(A)=1$, and $\mathcal{H}(B|A)$ is given by $\mathcal{H}(P_{01})$. The error in correlation between Alice and Bob is thus given by
$\mathcal{I}^E(A:B)=1-\mathcal{H}(P_{01})$.

Similarly, the error corresponding to case when the evesdropper Eve obtains odd parity while Alice 
tries to control Bob's state in even parity becomes
\begin{eqnarray}
Q_{01}(\alpha\pm\beta)&=&\sum_{m=0}^{\infty} |\langle 2m+1|\left(\alpha\pm\beta\right)\left(\frac{\eta}{|\eta|}\sin|\eta|-1\right)\rangle_A|^2 \nonumber \\
&=& \sinh\left(|\delta^{\prime}|^2\right)\,\exp\left(-\frac{|\delta^{\prime}|^2}{2}\right),
\end{eqnarray}
where $\delta^{\prime}=\left(\alpha\pm\beta\right)\left(\frac{\eta}{|\eta|}\sin|\eta|-1\right)$. Here, the average error is given by
$Q_{01}=\frac{1}{2}\,Q_{01}(\alpha+\beta)+\frac{1}{2}\,Q_{01}(\alpha-\beta)$, where we consider $P_{\alpha+\beta}=P_{\alpha-\beta}=\frac{1}{2}$ because randomness gives maximum error.
In this case, the error mutual information is given by
$\mathcal{I}^E(A:E)=1-\mathcal{H}(Q_{01})$.
 Thus, the bound on the error rate is given by~\cite{CsiszarKorner}
$r_e = \mathcal{I}(A:B)-\max_{Eve}\mathcal{I}^{E}(A:E)$,
where maximization is taken over all possible Eve's strategies. Here, the maximum occurs for 
$\eta=\pi/4$ and corresponding error rate becomes $0$. Such a result is analogous to the
key rate obtained to be $r=1$ for quantum continuous variable systems derived using the fine-grained
uncertainty relation~\cite{St_F_C}.

To summarize, in the present work we have derived a temporal steering criterion in classical optics.
Recently, quantum steering~\cite{epr, wise} has been recast as the control of a single quantum 
system at different times, with the formulation of a temporal steering scenario for discrete 
variable quantum systems~\cite{St_Tem_Dis, nonmarkov}. Here, by developing further the analogy
between quantum mechanics and classical wave optics emanating from the key feature of superposition
in both the theories~\cite{C-Op, mansuri, classent, classbell, priyanka}, we first formulate a 
fine-grained uncertainty relation in the realm of the latter. An
optimal~\cite{St_F_D, St_F_C} steering inequality thus follows using displaced parity operations
on a single mode coherent
state. Further, exploting the connection between temporal steering and the secret key rate of
the BB84 protocol, here we derive an analogous, and in principle, ideal key rate in classical 
optics. Note that though continuous variable quantum key generation using coherent states has
been proposed earlier~\cite{groshans}, the security therein is based on the quantum uncertainty 
principle. On the other hand, the security of key generation discussed here is based fundamentally 
upon the uncertainty relation (\ref{classuncert}) in classical wave optics. Besides exploring 
possible practical ramifications of this difference, it may also be interesting to perform further 
investigations on the ontological nature of a classical wave theory that permits steering and 
disallows perfect cloning, vis-a-vis its quantum counterpart~\cite{pbr}.

{\it Acknowledgements}: The authors acknowledge support from the project SR/S2/LOP-08/2013
of DST, India.

\end{document}